\documentclass[12pt]{article}

\usepackage{amsmath,amssymb,latexsym}
\usepackage[mathscr]{eucal}

\newcommand{\Ss}{\mathbb{S}}
\newcommand{\RR}{\mathbb{R}}
\newcommand{\LL}{\mathcal{L}}
\DeclareMathOperator{\tr}{tr}

\begin{document}

\begin{center}
    {\Large Knots as possible excitations of \\
    the quantum Yang-Mills fields }
\end{center}
\begin{center}
    L.~D.~Faddeev\\
    St.Petersburg Department\\
    of Steklov Mathematical Institute
\end{center}
    
\section{Dedication}
    It is a great honour and pleasure for me to participate in the
    conference, dedicated to 85 years celebration for Professor
    C.~N.~Yang.

    The influence of C.~N.~Yang on my own research is very strong.
    Two of my directions --- quantization of the Yang-Mills field
    and theory of solitons stem from his works. I am proud to remind,
    that the term ``Yang-Baxter equation'' was introduced by
    L.~Takhtajan and me and now covers extensive research in
    integrable models and quantum groups.

    In my talk I shall describe the subject, which to some extend
    connects solitons and Yang-Mills quantum field theory.
    As is reflected in the title, it is still not well established.
    However, I believe, that work on it will be continued in the future.

\section{Introduction}
    Quantum Yang-Mills theory
\cite{YM}
    is most probably the only viable relativistic field theory in
    4-dimensional space-time.
    The special property, leading to this conviction, is dimensional
    transmutation
\cite{Coleman}
    and related property of asymtotic freedom
\cite{GWPH}.
    However the problem of description of corresponding particle-like excitations
    is still not solved.
    The question, posed by W.~Pauli in 1954 during talk of C.~N.~Yang
    at Oppenheimer seminar at IAS
\cite{Op},
    waits for an answer for more than 50 years.
    In this talk I shall present a hypothetical scenario for this picture:
    particles of Yang-Mills field are knot-like solitons.

    The idea is based on another popular hypothese, according to which the
    confinement in QCD is effectuated by gluonic strings, connecting quarks.
    Thus a natural question is what happenes to these strings in the
    absence of quarks, i.~e. in the pure Yang-Mills theory.
    The strings should not disappear, they rather become closed,
    producing rings, links or knots.
    This idea was leading in my recent activity in collaboration
    with Antti Niemi.

    Our approach is based on a soliton model, which I proposed in the
    mid-70ties in the wake of interest to the soliton mechanism for
    particle-like excitations.
    My proposal was mentioned in several talks, partly refered to in
\cite{FNN}.

    The model is a kind of nonlinear
$\sigma$-model with nonlinear field
$ n(x) $
    taking values in the two-dimensional sphere
$ \Ss^{2} $.
    It does not allow complete separation of variables, so practical
    research was to wait until mid-90ties when computers strong enough
    became available.
    It was Antti Niemi, who was first to sacrifice himself for complicated
    numerical work with the great help of supercomputer center at Helsinki.
    The first result, published in
\cite{FNN},
    attracted attention of two groups
\cite{Hiet}, \cite{Bat}.
    Their work revealed rich structure of knot-like solitons, confirming
    my expectations.
    Thus a candidate for dynamical model with knot-like excitations
    was found.
    Next step was to find a place for this field among the
    dynamical variables of the Yang-Mills field theory.

    We developed consequentively two approaches for this.
    The first one was based on the proposal of Y.~M.~Cho
\cite{Cho}
    to construct kind of the magnetic monopole connection,
    described by means of the
$ n $-field
\cite{NF}.
    This approach is still discussed by several groups
\cite{Kondo}--\cite{Chox}.
    In fact Cho connection was found before in
\cite{Ge}.
    However, now we do not consider this approach as promising anymore
    and in the beginnning of new century developed another one.
    The short announcement
\cite{FN}
    was developed in a detailed paper
\cite{FNnp}.
    In this talk I shall briefly describe our way to this proposal
    and give its exposition.
    I shall begin with the description of the
$ \sigma $-model,
    then propose its application in the condenced matter theory and
    finally explain our approach to the Yang-Mills theory.

\section{Nonlinear $\sigma$-model}
    The field variable is
$ n $-field --- a unit vector
\begin{equation*}
    \vec{n} = (n_{1},n_{2},n_{3}) , \quad \sum n_{i}^{2} = 1 .
\end{equation*}
    In other words the target is a sphere
$ \Ss^{2} $.
    For static configurations the space variables ran through
$ \RR^{3} $.
    Boundary condition
\begin{equation*}
    n|_{\infty} = (0,0,1)
\end{equation*}
    compactifies
$ \RR^{3} $ to
$ \Ss^{3} $, so
$ n $-field realizes the map
\begin{equation*}
    n: \Ss^{3} \to \Ss^{2} .
\end{equation*}
    Such maps are classified by means of the topological charge,
    called Hopf invariant, which is more exotic in comparison with
    more usual degree of map, used when space and target have the
    same dimension.

    To describe this topological charge consider the preimage of the volume
    form on
$ \Ss^{2} $ --- 2 form on
$ \RR^{3} $ (or $\Ss^{3}$)
\begin{equation*}
    H = H_{ik} dx^{i} \wedge dx^{k}
\end{equation*}
    with
\begin{equation*}
    H_{ik} = (\partial_{i} \vec{n} \times \partial_{i} \vec{n}, \vec{n})
	= \epsilon_{abc} \partial_{i} n^{a} \partial_{k} n^{b} n^{c} ,
\end{equation*}
    which is exact
\begin{equation*}
    H = dC .
\end{equation*}
    Then Chern-Simons integral
\begin{equation*}
    Q = \frac{1}{4\pi} \int_{\RR^{3}} H \wedge C
\end{equation*}
    acquires only integer values and is called Hopf invariant.

    The formulas above have natural interpretation in terms of magnetic field.
    Indeed, the Poincare dual of 
$ H_{ik} $
\begin{equation*}
    B_{i} = \frac{1}{2} \epsilon_{ikj} H_{kj}
\end{equation*}
    is divergenceless
\begin{equation*}
    \partial_{i} B_{i} = 0
\end{equation*}
    and can be taken as a description of magnetic field.
    The preimage of a point on
$ \Ss^{2} $
    is a closed contour, describing a line of force of this field.
    The Hopf invariant is an intersection number of any two such lines.

    It is instructive to mention, that
$ n $-filed gives a way to describe the magnetic field alternative to one
    based on the the vector potential. In particular the configuration
\begin{equation*}
    \vec{n} = \frac{\vec{x}}{|x|}
\end{equation*}
    describes the magnetic monopole without annoying Dirac string.

    There are two natural functional, which can be used to introduce
    the energy. The first is the traditional
$ \sigma $-model
    hamiltonian
\begin{equation*}
    E_{1} = \int_{\RR^{3}} \bigl(\partial n\bigr)^{2} d^{3}x .
\end{equation*}
    The second is the Maxwell energy of magnetic field
\begin{equation*}
    E_{2} = \int_{\RR^{3}} \bigl(H_{ik}\bigr)^{2} d^{3} x .
\end{equation*}
    Functional
$ E_{1} $
    is quadratic in the derivatives of 
$ n $-field
    and 
$ E_{2} $
    is quartic in them. Thus they have opposite reaction to scaling
$ x \to \lambda x $
\begin{equation*}
    E_{1} \to \lambda E_{1} , \quad E_{2} = \frac{1}{\lambda} E_{2} ,
\end{equation*}
    which is reflected in their different dimensions
\begin{equation*}
    [E_{1}] = [L] , \quad [E_{2}] = [L]^{-1} .
\end{equation*}
    We take for the energy their linear combination
\begin{equation*}
    E = aE_{1} + b E_{2} ,
\end{equation*}
    where
\begin{equation*}
    [a] = [L]^{-2}
\end{equation*}
    and
$ b $
    is dimensionless. Derric theorem -- the well known obstruction for the
    existence of localized finite energy solutions (solitons) -- does not
    apply here.

    The estimate
\begin{equation*}
    E \geq c |Q|^{3/4} ,
\end{equation*}
    found in
\cite{VK},
    supports the belief that such solutions do exist. Unfortunately
    the relevant mathematical theorem is not proved until now, so we are
    to refer to numerical evidence
\cite{Hiet}, \cite{Bat}.
    The picture of solutions looks as follows. The lowest energy
$ Q=1 $
    soliton is axial symmetric; it is concentrated along the circle
$ n_{3} = -1 $;
    the magnetic surfaces (preimages of lines $n_{3} = \text{const} $)
    are toroidal, wrapped once by by magnetic lines of force.
    For
$ Q=4 $
    minimal solution is a link and for
$ Q=7 $
    it is trefoil. Beautiful computer movies, illustrating the calculations
    based on the descent method, can be found in
\cite{Hietm}.

    There is a superficial analogy of the
$ \sigma $-model
    with the Skyrme model
\cite{Sk}
    for the principal chiral field
$ g(x) $
    with values in the manifold of compact Lie group
$ G $.
    Skyrme lagrangian is expressed via the Maurer-Cartan current
\begin{equation*}
    L_{\mu} = \partial_{\mu} g g^{-1}
\end{equation*}
    as follows
\begin{equation*}
    \LL = a \tr L_{\mu}^{2} + b \tr [L_{\mu}, L_{\nu}]^{2} ,
\end{equation*}
    which also contains terms quadratic and quartic in derivatives of
$ g $.
    Corresponding topological charge
\begin{equation*}
    Q = \int \tr [L_{i},L_{k}] L_{j} \epsilon_{ikj} d^{3} x
\end{equation*}
    coinsides with the degree of map for
$ G = SU(2) $.
    There is an estimate for static Hamiltonian
\begin{equation*}
    E \geq c |Q| .
\end{equation*}
    The minimal excitation for
$ Q=1 $
    is spherically symmetric and concentrated around a point.

    So there are two important differences between two models.
    First, the excitations of Skyrme model are point like, whereas those
    for nonlinear
$ \sigma $-model are string-like.
    Second the term
$ E_{2} $
    has natural interpretation as Maxwell energy whereas the quartic term
    in the Skyrme model is rather artificial.

    This concludes the description of the nonlinear
$ \sigma $-model and I must turn to its applications.
    Before the main one to Yang-Mills field, I shall consider
    more simple example, developed together with Niemi and Babaev
\cite{Bab}.

\section{Two component Landau-Ginsburg-Gross-Pitaevsky equation}
    The equation from the title appears in the theory of
    superconductivity (LG) and Bose gas (GP). The main degree of freedom
    is a complex valued function
$ \psi(x) $ --- gap in the superconductivity or density in Bose-gas.
    Magnetic field is described by vector potential
$ A_{k}(x) $.
    There is a huge literature
    dedicated to the LGGP equation. Our contribution consists in using
    two components
$ \psi $
\begin{equation*}
    \psi = (\psi_{1}, \psi_{2}) ,
\end{equation*}
    corresponding to a mixture of two materials.

    The energy in the appropriate units is written as
\begin{equation*}
    E = \sum_{\alpha=1}^{2} |\nabla_{i} \psi_{\alpha}|^{2} + F_{ik}^{2}
	+ v (\psi) ,
\end{equation*}
    where
\begin{equation*}
    \nabla_{i} \psi = \partial_{i} \psi + iA_{i} \psi
\end{equation*}
    and
\begin{equation*}
    F_{ik} = \partial_{i} A_{k} - \partial_{k} A_{i} .
\end{equation*}
    The functional
$ E $
    is invariant with respect to the abelian gauge transformation
\begin{equation*}
    A_{i} \to A_{i} + \partial_{i} \lambda , \quad
	\psi_{\alpha} \to e^{-i\lambda} \psi_{\alpha}
\end{equation*}
    with an arbitrary real function
$ \lambda $.
    In the case of one component
$ \psi $
    the change of variables
\begin{equation*}
    \psi = \rho e^{i\theta} , \quad A_{k} = B_{k} + \frac{1}{\rho^{2}} J_{k} ,
\end{equation*}
    where
\begin{equation*}
    J_{k} = \frac{1}{2i} \bigl(\overline{\psi}\partial_{k}\psi 
	    - \overline{\partial_{k}\psi} \psi \bigr) ,
\end{equation*}
    transforms
$ E $
    to the gauge invariant form
\begin{equation*}
    E = (\partial \rho)^{2} + \rho^{2} B^{2} +
	(\partial_{i}B - \partial_{k}B)^{2} + v(\rho) ,
\end{equation*}
    eliminating phase
$ \theta $
    and leaving gauge invariant density
$ \rho $
    and supercurrent
$ B $.
    The potential
$ v(\rho) $
    is supposed to produce the nonzero mean value for
$ \rho $
\begin{equation*}
    <\rho> = \Lambda ,
\end{equation*}
    vector field
$ B $
    becomes massive (Meissner effect with finite penetration length).

    In the case of two components
$ \psi_{\alpha}, \alpha=1,2 $
    the analogous change of variables, proposed in
\cite{Bab},
    looks as follows
\begin{gather*}
    \rho^{2} = |\psi_{1}|^{2} + |\psi_{2}|^{2} , \\
    \vec{n} = \frac{1}{\rho^{2}} (\bar{\psi}_{1}, \bar{\psi}_{2})
	\vec{\tau}
    \begin{pmatrix}
	\psi_{1} \\
	\psi_{2}
    \end{pmatrix} \\
    A_{k} = B_{k} + \frac{1}{\rho^{2}} J_{k} \\
    J_{k} = \frac{1}{2i} \sum_{\alpha} \bigl(
	\bar{\psi}_{\alpha} \partial_{k} \psi_{\alpha}
	- \partial_{k} \bar{\psi}_{\alpha} \psi_{\alpha} \bigr) .
\end{gather*}
    Here
$ \vec{\tau} = (\tau_{1},\tau_{2},\tau_{3}) $
    is set of Pauli matrices
\begin{equation*}
    \tau_{1} = \begin{pmatrix}
    0 & 1 \\
    1 & 0
    \end{pmatrix} , \quad
    \tau_{2} = \begin{pmatrix}
    0 & -i \\
    i & 0
    \end{pmatrix} , \quad
    \tau_{3} = \begin{pmatrix}
    1 & 0 \\
    0 & -1
    \end{pmatrix} .
\end{equation*}
    Variables
$ \rho $, $ B $ and $ n $ are gauge invariant.
    Thus the difference with the case of one component is appearence of the
$ n $-field.

    The energy in new variables looks as follows
\begin{equation*}
    E = (\partial\rho)^{2} + \rho^{2} (\partial n)^{2}
	+ \bigl(\partial_{i} B_{k} -\partial_{k} B_{i} +H_{ik}\bigr)^{2}
	+ \rho^{2} B^{2} + v(\rho)
\end{equation*}
    and contains both ingredients of the nonlinear
$ \sigma$-model from section 1.
    If due to the Meissner effect massive vector field
$ B $
    vanishes in the bulk, only
$ n $-field remains there and should produce knot-like excitations.
    This is our main prediction and we wait for the relevant experimental work.

    I want to stress the difference of our excitations with Abrikosov
    vortices. Our closed strings have finite energy in 3-dimensional
    bulk, whereas Abrikosov vortices are two-dimensional.
    Moreover, the corresponding topological
    charges are distinct --- Hopf invariant in our case and degree of map
$ \Ss^{1} \to \Ss^{1} $
    in the case of Abrikosov vortices.

    Now it is time to turn to the main subject --- Yang-Mills field.

\section{$SU(2)$ Yang-Mills theory}
    The field variables are 3 vector fields
$ A_{\mu}^{a}, a=1,2,3 $,
    describing connection in the fiber bundle
$ M_{4}\times SU(2) $,
    where
$ M_{4} $
    is a space-time, which for definiteness we shall take as euclidean
$ \RR^{4} $.
    Let
$ \tau^{a} $
    be Pauli matrices and
\begin{equation*}
    A_{\mu} = A_{\mu}^{a} \tau^{a} .
\end{equation*}
    The gauge tranformation is given by
\begin{equation*}
    A_{\mu} \to g A_{\mu} g^{-1} + \partial_{\mu} g g^{-1}
\end{equation*}
    with arbitrary
$ 2\times 2 $ unitary matrix
$ g $.
    The curvature (field strength)
$ F_{\mu\nu} $
\begin{equation*}
    F_{\mu\nu} = \partial_{\mu} A_{\nu} - \partial_{\nu} A_{\mu}
	+ [A_{\mu},A_{\nu}]
\end{equation*}
    transforms homogeneously
\begin{equation*}
    F_{\mu\nu} \to g F_{\mu\nu} g^{-1}
\end{equation*}
    and Largangian
\begin{equation*}
    \LL_{\text{YM}} = \frac{1}{4} \tr (F_{\mu\nu})^{2}
\end{equation*}
    is gauge invariant.

    The maximal abelian partial gauge fixing (MAG), which we shall use,
    put restriction on the offdiagonal components
$ A_{\mu}^{1} $ and $ A_{\mu}^{2} $.
    We shall use the complex combination
\begin{equation*}
    B_{\mu} = A_{\mu}^{1} + iA_{\mu}^{2}
\end{equation*}
    and MAG condition looks as follows
\begin{equation*}
    \nabla_{\mu} B_{\mu} = 0,
\end{equation*}
    where
\begin{equation*}
    \nabla_{\mu} = \partial_{\mu} +iA_{\mu} , \quad A_{\mu} = A_{\mu}^{3} .
\end{equation*}
    The fact, that we use a distinguished (diagonal) direction in the
    charge space is not essential, see
\cite{FNnp}
    for details.

    The remaining gauge freedom is the abelian one
\begin{equation*}
    B_{\mu} \to e^{-i\lambda} B_{\mu} , \quad 
	A_{\mu} \to A_{\mu} + \partial_{\mu} \lambda .
\end{equation*}

    MAG condition can be realized by adding 
    the quadratic form
$ \frac{1}{2}|\nabla_{\mu}B_{\mu}|^{2} $ to
$ \LL_{\text{YM}} $,
    leading to
\begin{multline*}
    \LL_{\text{MAG}} = \LL_{\text{YM}} 
	+ \frac{1}{2} |\nabla_{\mu} B_{\mu}|^{2} =\\
    = \frac{1}{2} |\nabla_{\mu}B_{\nu}|^{2}
	+ \frac{1}{4} (F_{\mu\nu}+H_{\mu\nu})^{2}
	+ \frac{1}{2} F_{\mu\nu} H_{\mu\nu} ,
\end{multline*}
    where
\begin{equation*}
    F_{\mu\nu} = \partial_{\mu}A_{\nu}-\partial_{\nu}A_{\mu} , \quad
	H_{\mu\nu} = \frac{1}{2i} (\bar{B}_{\mu}B_{\nu}-\bar{B}_{\nu}B_{\mu}).
\end{equation*}
    The last term appears after the integration by parts, used to
    eliminate the unwanted term
$ \bar{\nabla}_{\mu}\bar{B}_{\nu} \nabla_{\nu} B_{\mu} $.

    Now I come to the main trick.
    Observe, that two vector fields
$ A_{\mu}^{1} $ and
$ A_{\mu}^{2} $
    define 2-plane in
$ M_{4} $.
    Let us parametrize this 2-plane by the orthogonal zweibein
$ e_{\mu} $
\begin{gather*}
    e_{\mu} = e_{\mu}^{1} + i e_{\mu}^{2} \\
    \bar{e}_{\mu} e_{\mu} = 1 , \quad e_{\mu}^{2} = \bar{e}_{\mu}^{2} = 0
\end{gather*}
    and express
$ B_{\mu} $
    as
\begin{equation*}
    B_{\mu} = \psi_{1} e_{\mu} + \psi_{2} \bar{e}_{\mu} ,
\end{equation*}
    introducing two complex coefficients
$ \psi_{1} $ and
$ \psi_{2} $.
    Altogether the set
$ e_{\mu}, \psi_{1}, \psi_{2} $
    contains 9 real functions and
$ B_{\mu} $
    has only 8 real components.
    The discrepancy is resolved by comment, that expression for
$ B_{\mu} $
    is invariant with respect to the abelian gauge transformation
\begin{equation*}
    e_{\mu} \to e^{i\omega}e_{\mu} , \quad \psi_{1} \to e^{-i\omega} \psi_{1},
	\quad \psi_{2} \to e^{i\omega} \psi_{2} .
\end{equation*}
    Corresponding
$ U(1) $
    connection is given by
\begin{equation*}
    \Gamma = \frac{1}{i} (\bar{e}_{\nu} \partial_{\mu} e_{\nu}) , \quad
	\Gamma_{\mu} \to \Gamma_{\mu} + \partial_{\mu} \omega .
\end{equation*}
    Now having
$ \psi_{1} $, $ \psi_{2} $
    we can repeat trick from section 2, introducing
$ n $-field.
    However in our case we can do more. Indeed, the combination
$ H_{\mu\nu} $,
    entering
$ \LL_{\text{MAG}} $,
    can be written as
\begin{equation*}
    H_{\mu\nu} = \rho^{2} n_{3}^{2} g_{\mu\nu}
\end{equation*}
    with
\begin{equation*}
    \rho^{2} n_{3}^{2} = |\psi_{1}|^{2} - |\psi_{2}|^{2} , \quad
	g_{\mu\nu} = \frac{1}{2i} (\bar{e}_{\mu} e_{\nu} 
	    - \bar{e}_{\nu}e_{\mu}).
\end{equation*}
    Putting
\begin{equation*}
    p_{i} = g_{0i} , \quad q_{i} = \frac{1}{2} \epsilon_{ijk} g_{jk}
\end{equation*}
    we get two vectors
$ p_{i} $, $ q_{i} $
    satisfying conditions
\begin{equation*}
    p^{2} + q^{2} = 1 , \quad (p,q) = 0 ,
\end{equation*}
    thus defining two spheres
$ \Ss^{2} $.
    Indeed, what we get here is a particular parametrization of the
    Grassmanian
$ G(4,2) $.
    In static case
$ p_{i} $
    disappears and we are left with one unit 3-vector
$ q $,
    which evidently could be used to introduce the magnetic monopoles.

    Now we can put the new variables into
$ \LL_{\text{MAG}} $.
    All details are to be found in
\cite{FNnp}.
    Here I shall write explicitely the static energy
\begin{multline*}
    E = (\partial_{i} \rho)^{2} + \rho^{2} \bigl((\nabla_{k}n)^{2}
	+ (\partial_{k}q)^{2}\bigr) + \rho^{2} C_{k}^{2} +\\
    + \frac{1}{4}\bigl((\partial_{i}n \times \partial_{k}n,n) +
	(\partial_{i}q \times \partial_{k}q,q) + 
	2 H_{ik} +\partial_{i}C_{k} -\partial_{k}C_{i}
	\bigr)^{2} - \frac{3}{4} \rho^{4} n_{3}^{2} ,
\end{multline*}
    where
$ C $
    is supercurrent
\begin{equation*}
    C_{k} = A_{k} + \frac{1}{2\rho^{2}} \bigl(
	\bar{\psi}_{1}(\partial_{k}+iA_{k}+i\Gamma_{k})\psi_{1} +
	\bar{\psi}_{2}(\partial_{k}+iA_{k}-i\Gamma_{k})\psi_{2}
	- \text{c.c.}\bigr)
\end{equation*}
    and
\begin{equation*}
    \nabla_{k} n^{a} = \partial_{k} n^{a} + \epsilon^{ab3} \Gamma_{k} n^{b} .
\end{equation*}
    We see, that the structure of nonlinear
$ \sigma $-model appears twice --- via fields
$ n $ and $ q $.
    We can interprete it as a new manifestation of electromagnetic duality
    in the nonabelian Yang-Mills theory.

    The expression for
$ E $
    can be taken as a point of departure for speculations on the
    knot-like excitations for the
$ SU(2) $
    Yang-Mills field. The corresponding transformation for
$ SU(3) $
    case, done in
\cite{FB},
    is more complicated due to difference of rank and number of roots.

    I want to stress, that by no means I propose to use the new variables
    to make a change of variables in the functional integral.
    Rather they should be put into the renormalized effective action,
    which should be found in the background field formalism.
    The variant of this method, where the background field is not
    classical, but is a solution of the quantum modified equation
    of motion is given in
\cite{back}.
    In the course of renormalization this effective action should
    experience the dimensional transmutation. We still do not know,
    how it happens, so we can only use speculations.
    The main hope, that in this way the mean value of
$ \rho^{2} $
    will appear. At the same time the classical lagrangian should be
    main part of the effective action, as it represents the only possible
    local gauge invariant functional of dimension -4. The condensate of
$ <\rho^{2}> $
    of dimension -2 is the subject of many papers in the last years
(see e.~g. \cite{Z}, \cite{V}).
    Thus all this makes the picture of string-like excitations for the
    Yang-Mills field more feasible. However the real work
    only begins here. I hope, that this subject will take fancy of some more
    young researchers.

\end{document}